\documentclass[12pt]{article}

\usepackage{amsmath,amssymb,graphicx} 
\usepackage{cite}

\newcommand{\beq}{\begin{eqnarray}}
\newcommand{\eeq}{\end{eqnarray}}

\newcommand{\centeron}[2]{{\setbox0=\hbox{#1}\setbox1=\hbox{#2}\ifdim

\wd1>\wd0\kern.5\wd1\kern-.5\wd0\fi
\copy0

\kern-.5\wd0\kern-.5\wd1\copy1\ifdim\wd0>\wd1
                                       \kern.5\wd0\kern-.5\wd1\fi}}
\newcommand{\ltap}{\>\centeron{\raise.35ex\hbox{$<$}}
                               {\lower.65ex\hbox{$\sim$}}\>}
\newcommand{\gtap}{\>\centeron{\raise.35ex\hbox{$>$}}
                               {\lower.65ex\hbox{$\sim$}}\>}

\newcommand\ZZ{\hbox{\zfont Z\kern-.4emZ}}
\font\zfont = cmss10 
\newcommand{\sfrac}[2]{{\textstyle\frac{#1}{#2}}}

\begin{document}
\begin{titlepage}
\begin{flushright}
{\tt hep-ph/0604218}
\end{flushright}

\vskip.5cm
\begin{center}
{\huge \bf Field Theory on\\
\vspace{.2cm}
Multi-throat Backgrounds
}

\vskip.1cm
\end{center}
\vskip0.2cm

\begin{center}
{\bf
{Giacomo Cacciapaglia}$^{a}$, {Csaba Cs\'aki}$^{a}$,
{Christophe Grojean}$^{b,c}$\\
{\rm and}
{John Terning}$^{d}$}
\end{center}
\vskip 8pt

\begin{center}
$^{a}$ {\it Institute for High Energy Phenomenology\\
Newman Laboratory of Elementary Particle Physics\\
Cornell University, Ithaca, NY 14853, USA } \\
\vspace*{0.1cm}
$^{b}$~{\it CERN Physics Department, Theory Division, CH-1211 Geneva 23, Switzerland}\\
\vspace*{0.1cm}
$^{c}$ {\it Service de Physique Th\'eorique,
CEA Saclay, F91191 Gif--sur--Yvette, France} \\
\vspace*{0.1cm}
$^{d}$ {\it
Department of Physics, University of California, Davis, CA
95616.} \\
\vspace*{0.3cm}
{\tt  cacciapa@mail.lns.cornell.edu, csaki@lepp.cornell.edu,
grojean@spht.saclay.cea.fr,  terning@physics.ucdavis.edu}
\end{center}

\vglue 0.3truecm

\begin{abstract}
\vskip 3pt
\noindent
We consider extra dimensional field theory descriptions of  backgrounds with $N$~different throats
where
one of the extra dimensions in each throat is much larger than the others. Such backgrounds
can be described by  field theory on $N$~5D warped spaces which intersect on a ultraviolet (UV) brane.
Given a field that propagates in all $N$~throats there are $N$~boundary conditions on the UV
brane (which are determined by the effective Lagrangian on the UV brane) in addition to the
boundary conditions on the $N$~infrared branes.  We derive a general set of UV boundary conditions
and give  examples of how they are applied to particular situations.  Three simple example applications are given:
 in the first the number of families is determined by the number of throats and the
SUSY flavor problem is solved via an $S_3$ symmetry of the throats;
in the second  we embed this scenario in  a SUSY GUT with  a solution of the doublet-triplet splitting problem based on the product group approach; while in the final example
we show a simple geometric implementation of a SUSY trinification model on three throats.

\end{abstract}

\end{titlepage}

\newpage


\section{Introduction}
\label{sec:intro}
\setcounter{equation}{0}
\setcounter{footnote}{0}

\begin{figure}[htb]
\begin{center}
\includegraphics[width=6cm]{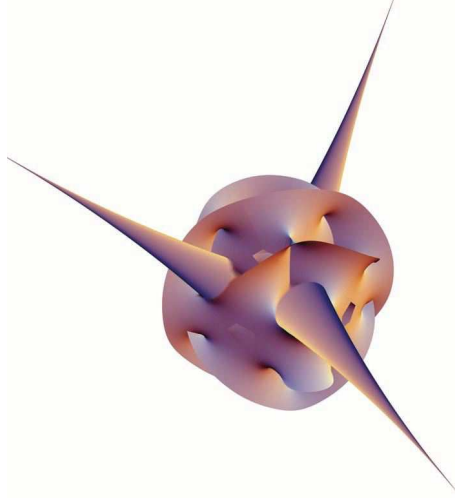}
\end{center}
\caption{A generic multi-throat geometry. Multiple throats are hanging out from the head which is a compact Calabi-Yau manifold. In our field theory approach, each throat is modelled by a single warped extra dimension. The throats intersect at a point, the UV brane. The UV physics of the head is described by some brane localized operators connecting the various fields living in the individual throats.} \label{fig:throats}
\end{figure}

Theories with extra dimensions have found many new applications over the past
ten years. Large and warped extra dimensions have been proposed
to alleviate the hierarchy problem~\cite{ADD,RS}, small extra dimensions
can give rise to novel approaches to GUT model building~\cite{GUTorbifold},
universal extra dimensions can give new dark matter candidates~\cite{KKDM},
and even new mechanisms for electroweak symmetry breaking can be
found using extra dimensions~\cite{A5Higgs, CGMPT, CGPT}. In most of these models, only the simplest
possible topologies of the extra dimensional space have been assumed: a single
(or sometimes two) extra dimensions compactified on a circle (or torus), or a
very simple orbifold like $S^1/Z_2$ or $T^2/Z_2$. This severely
limits the possible configurations available for model building. For example
in the case of warped extra dimensions it is usually assumed that there
is a single warp factor in the theory, and that one needs to introduce
complicated intersecting brane configurations in order to get a warped
model in more than one extra dimension. However, it has been found in string
theory~\cite{StringThroats} that a generic type of situation arising from flux compactifications
leads to geometries with multiple ``throats'' hanging out from the ``head'',
which is a compact Calabi--Yau manifold (see fig.~\ref{fig:throats}).
Such compactifications with multiple
throats could have many advantages over the more commonly considered
ones. It is relatively simple to introduce multiple scales into the theory,
it allows for clearly separate fields  without having to
introduce multiple extra dimensions, etc. The purpose of this paper is
to give a simple field theory description of such multi-throat
string compactification set-ups, and to show some possible simple
applications for extra dimensional model building that would not be
easily achieved without such set-ups (see also~\cite{NemanjaEtal} for previous phenomenological applications of multi-throat set-ups).

\section{Boundary Conditions}
\label{sec:scalar}
\begin{figure}[htb]
\begin{center}
\includegraphics[width=11cm]{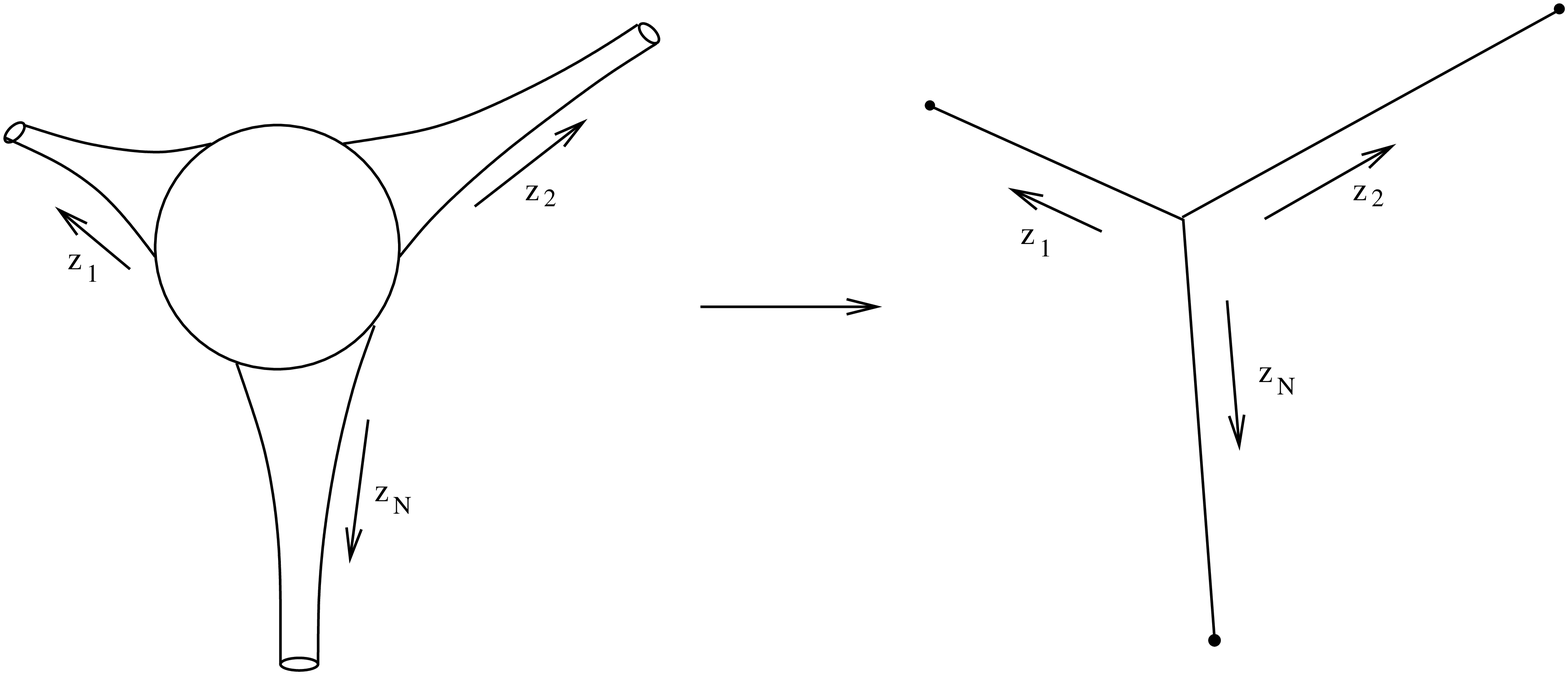}
\end{center}
\caption{Illustration of the 
limit of the generic multi-throat geometry considered
in this paper.}
\label{fig:limit}
\end{figure}

\setcounter{equation}{0}
\setcounter{footnote}{0}
In the field theory limit that we are considering we imagine that
the transverse sizes of the throats are all negligible, and also
shrink the ``head'' to a single point, the UV brane. Thus we end up
with $N$~5D bulks parametrized by the coordinates $z_1,\ldots , z_N$
(see fig.~\ref{fig:limit}). We
should again stress that these are {\it not} coordinates of $N$~extra dimensions,
but rather $N$~single extra dimensions intersecting at a point in space,
thus identifying the points\footnote{For a flat extra dimension, we can always chose the UV brane in each throat to be located at $z_i=0$. In a warped extra dimension where translational invariance along the extra dimension is lost, we cannot simply shift the origin of the coordinates $z_i$ to bring the UV brane at $z_i=0$. The discussion presented here is easily generalized to more general (warped) set-ups.} $z_1=0$ with $z_2=0, \ldots ,z_N=0$.
This point is what we will be calling the UV brane. The other~$N$
(distinct) endpoints of
the throats corresponding to $z_1=R'_1$, $z_2=R'_2,\ldots , z_N=R'_N$ will
correspond to $N$~separate IR branes, thus yielding $N$~separate scales in the
theory. This set-up is a generalization to multi-throats of our construction~\cite{Cacciapaglia:2005pa} where two throats have been introduced to separate the top and bottom quarks from the light fermions.

On this geometry, let us consider a single scalar field $\phi$
which propagates in all of the throats. We start our analysis of the
modes of this field by introducing $N$~5D bulk scalar fields $\phi_i$
each of which will live in just one of the throats that
intersect on the single UV brane. The bulk actions are given by\footnote{Throughout this paper,we'll use a mostly minus signature and $M,N \ldots =0,1,2,3,5$ denote bulk coordinates while $\mu,\nu \ldots=0,1,2,3$ denote brane coordinates.}
\beq
{\mathcal L}_{bulk}=
\sum_i
\sqrt{g_{(i)}} \left(
\frac{1}{2} g_{(i)}^{MN}\partial_M\phi_i \partial_N \phi_i-V_i(\phi_i) \right)~,
\eeq
where $g_{i}$ is the metric in the $i$-th throat.
In order for the picture to be consistent, they have to match on the UV brane: $g_{(1)|z_1=UV} = \dots = g_{(N)|z_N=UV} = g_{(UV)}$.
At this point, the $N$~fields $\phi_i$ are still totally independent.
In order to make it a single field propagating over the whole geometry
we need to connect these fields at the UV brane.
For simplicity we will first consider only the lowest dimension operators on the UV brane (warp factor dependencies are included in the mass terms):
\beq
{\mathcal L}_{UV}=
\left( -\frac{1}{2} \sum_{i,j=1}^N m_{ij} \phi_i \phi_j \right)_{| z_i=UV}~.
\eeq
The variation of the bulk actions and the UV brane action gives the following $N$~boundary conditions (BCs):
\beq
0= \partial_{z_i} \left( g_{(i)}^{55} \sqrt{g_{(i)}} \phi_i \right) + \sum_{j=1}^N m_{ij} \phi_j~.
\label{boundcond}
\eeq
UV brane kinetic terms can be included in a straightforward fashion, leading to $\partial^\mu\partial_\mu \phi_i$  terms in the BC.  Including interaction terms on the UV brane will lead to nonlinear terms in the BCs.
Note, that (\ref{boundcond}) provides exactly the right number of BC's to be able to fully determine
the system. We started out with $N$~scalar fields each of which satisfy a second order differential
equation, so one needs $2N$ BC's to solve the system. $N$~of these are provided by the BC's at the IR branes,
while further $N$~come from (\ref{boundcond}) at the UV brane.

\begin{figure}[htb]
\begin{center}
\includegraphics[width=5cm]{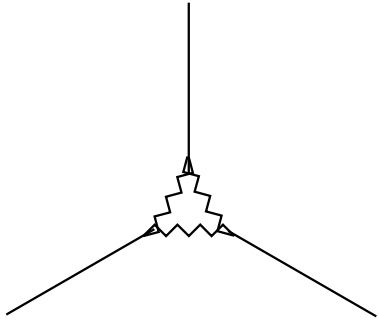}
\end{center}
\caption{A mechanical analogy: our procedure of generating the BC's on the UV brane can be thought of as
vibrating rods connected by springs at one of their ends.}
\label{analogy}
\end{figure}

A particular choice of mass terms on the UV brane leads to a very simple set of BCs.
If the mass terms have the form (with the convention that $\phi_{N+1}=\phi_1$)
\beq
    \label{simple}
{\mathcal L}_{UV}
=
- \sum_{i=1}^{N} k_i\, m \, (\phi_i -\phi_{i+1})^2_{| z_i=UV}
\eeq
then the BCs are
\begin{equation}
\partial_{z_i}  \left( g_{(i)}^{55} \sqrt{g_{(i)}} \phi_i \right) +2 k_i m (\phi_i -\phi_{i+1})=0.
\end{equation}
In the large $m$ limit these equations will just simplify to the $N-1$ independent equations (independent
of the $k_i$'s):
\beq
\phi_i &=&\phi_{i+1} \hspace{1cm} \mathrm{for}\, i=1\ldots N-1
\label{simpleuvbc0}
\eeq
while the sum of these equations will give the remaining BC
\beq
0&=& \sum_{i=1}^N  \sfrac{1}{k_i} \partial_{z_i}   \left( g_{(i)}^{55} \sqrt{g_{(i)}} \phi_i \right)
\label{simpleuvbc}
\eeq
A mechanical analogy for this system is given by $N$~vibrating rods, which are connected at their ends with springs to each other, with the boundary mass $m$ corresponding to the spring constant (see Fig.~\ref{analogy}).

In a more general case, the structure of the spectrum, in particular the number of zero modes and their localization in the throats, depends on the BC on the UV brane, effectively described by the localized mass terms.
The general set of BCs in eq.~(\ref{boundcond}) can be rewritten as:
\beq
0= \partial_{z_i}  \phi_i  + \sum_{j=1}^N \tilde m_{ij} \phi_j\,,
\eeq
where
\beq
\tilde m_{ij} = \left( g_{(UV)}^{55} \sqrt{g_{(UV)}} \right)^{-1}  \big( \partial_{z_i} \left( g_{(i)}^{55} \sqrt{g_{(i)}} \right) \delta_{ij} + m_{ij} \big)_{| z_i = UV}\,
\eeq
(we have used here the continuity of the metric on the UV brane, and simply redefined the mass term in order to include the derivative of the metric factor).
Let's assume for simplicity that bulk mass terms are absent and the BC's on the IR branes are Neuman: in this case the zero modes are characterized by flat wave functions $\phi_i = \mbox{const}$.
Therefore, there is a zero mode for every BC involving only derivatives: after diagonalizing the mass matrix $\tilde m$, we can see that the number of zero modes is equal to the number of zero eigenvalues ($N-$rank $\tilde m$).
The eigenstates also determine where the zero modes live.
Take $v_i^{(a)}$ to be the components of $a-$th eigenvector with eigenvalue $0$ (i.e. $a$ runs  from  $1$ to $N-$rank $\tilde m$).  Then the solutions for  the constant zero modes are
\beq
\phi_i^{(a)} \propto v_i^{(a)}~.
\eeq
If $v_k$ is smaller that the other coefficients, the component of the zero mode living in the $k$-th throat is suppressed with respect to the others, so that the zero mode is repelled from the $k$-th throat.
On the other hand, if $v_k$ is larger than the other $v$'s, than the zero mode lives preferably in the $k$-th throat.
In this way, just playing with the BCs we can control where a scalar zero mode is living.
  For the more  general case with non-constant  zero modes bulk solutions (but still assuming that the warp factors and bulk masses are the same in each throat), the actual zero modes correspond to eigenvectors of $\tilde m_{ij}$ whose eigenvalues are the negative of the slope of the zero mode bulk solution.  It is straightforward to generalize this to the case where the bulk zero mode solutions are different in each throat.

It is easy to check that for a {\it flat} bulk with the general UV BCs (\ref{boundcond}) that the standard
reflection/transmission problem works correctly.  We can for instance check it explicitly in the non-trivial case of three throats connected at a UV brane by general BCs of the form~(\ref{boundcond}).
Taking IR BCs such that on one bulk
there is an incoming (towards the UV brane) plane wave and a reflected plane wave, while in the other bulks there are transmitted (moving away from the UV brane) plane waves, the wave functions of the three scalar fields will be of the form (the UV BCs require the waves to have the same frequency in the different throats and, assuming the same bulk mass in the different throats, the bulk equations of motion also impose the wave vectors to have the same strength)
\beq
& \phi_1 (x,z_1) = e^{i(\omega t - k z_1)} + r \, e^{i(\omega t + k z_1)} \\
& \phi_2 (x,z_2) = \alpha_2 e^{i(\omega t + k z_2)}  \\
& \phi_3 (x,z_3) = \alpha_3 e^{i(\omega t + k z_3)}
\eeq
$r$ is the reflection coefficient in the first throat, while $\alpha_{2,3}$ are the transmission coefficients in the other throats.
The general BCs at the UV brane ($z_i=0$) are
\beq
\partial_{z_1} \phi_{1\, |z_1=0} + m_{11} \phi_{1\ |z_1=0} + m_{12} \phi_{2\ |z_2=0}
+ m_{13} \phi_{3\, |z_3=0} = 0 \nonumber \\
\partial_{z_2} \phi_{2\, |z_2=0} + m_{12} \phi_{1\ |z_1=0} + m_{22} \phi_{2\ |z_2=0}
+ m_{23} \phi_{3\, |z_3=0} = 0 \nonumber \\
\partial_{z_3} \phi_{3\, |z_3=0} + m_{13} \phi_{1\ |z_1=0} +  m_{23}  \phi_{2\ |z_2=0}
+ m_{33} \phi_{3\, |z_3=0} = 0
\label{general}
\eeq
These three equations can be solved to obtain the reflection and transmissions coefficients in terms of the UV brane masses, $m_{ij}$. For example, in the simplest case with the boundary terms
as in (\ref{simple}) and of three throats in the limit $m\to \infty$ the BC's will be
\beq
& \phi_{1\, |z_1=0}= \phi_{2\, |z_2=0}=\phi_{3\, |z_3=0}, \nonumber \\
& \partial_1 \phi_{1\, |z_1=0}+\partial_2 \phi_{2\, |z_2=0}+\partial_3 \phi_{3\, |z_3=0}=0.
\eeq
One can then explicitly calculate that
\beq
r=-\frac{1}{3}, \ \alpha_2=\frac{2}{3}, \ \alpha_3=\frac{2}{3} ,
\eeq
and thus
\beq
|r|^2+|\alpha_2|^2+|\alpha_3|^2 =1.
\eeq
This is a simple example of conservation of probability, that is the sum of the reflection coefficient and
the transmission coefficients into the different throats adds up to one:
\beq
R+\sum_{i=1}^N T_i = 1
\eeq
 One can show that this relation is also
satisfied for the general case with the BC's from (\ref{general}).
This simple result can be easily generalized to incoming waves on an arbitrary number of throats.

\subsection{Spectrum of KK modes}

The spectrum of KK modes in the `thin' limit of a multi-throat configuration
possesses some interesting features.   A free scalar field propagating on $N$~flat
throats joined at a single UV brane provides a simple case that shows many aspects
of the general situation.  Applying Neumann BCs at the IR ends
of each of the throats $\phi'(R'_i)=0$ and the simple Kirchhoff-type
BCs given in Eqs. (\ref{simpleuvbc0})--(\ref{simpleuvbc}) at the UV brane,
the solution in the $i$'th throat for the $n$'th KK wavefunction is given by
\beq
\phi_{i,n}(z_i) = \sqrt{2} C_n  {\cos [q_n (R'_i-z_i)] \over \cos (q_n R'_i) }
\eeq
where the KK mass-squareds are determined by the solutions for $q_n$ of the
eigenvalue equation
\beq
\sum_{i=1}^{N}  \tan(q_n R'_i) = 0 .
\label{keigen}
\eeq
The normalization constant $C_n$ is found to be
\beq
C_n = {1\over \sqrt{\sum_i R'_i \sec^2(q_n R'_i)}} .
\label{norm}
\eeq

If the $N$~throats are symmetric $R'_1=R'_2=\ldots R'_N\equiv R'$ then the
solutions to the KK momentum eigenvalues are simple
\beq
q_n = {n\pi \over R'}
\eeq
with normalized wavefunction in the $i$'th throat
\beq
\phi_{i,n}(z_i) = \sqrt{\frac{2}{N R'}}  \cos[q_n (R'-z_i)] .
\label{norm2}
\eeq
Notice that this normalization factor depends upon the total volume of the $N$~throats
but the mass gap to the first excited excited state above the zero mode is $1/R'$.
This leads to the possibility of large volume compactifications without the usual
associated light KK modes cf.\cite{Kaloper:2000jb}.

However if the throats are not symmetric, as we would generically expect, then the
spectrum of KK modes can be extremely complicated, even in this very simple flat-throat free-scalar case.  In particular, if the throat lengths $R'_i$ are not all rationally related then the spectrum of KK masses appears to be `chaotic' as can be seen by simple numerical investigation of the eigenvalue equation (\ref{keigen}).
Alternatively Dirichlet conditions at the UV brane can lead to the spectrum of $N$~decoupled throats.

\section{Fermions in multiple throats}
\label{sec:fermions}
\setcounter{equation}{0}
\setcounter{footnote}{0}

Fermions in extra dimensions always require special attention, since chiral fermions
appear only after imposing appropriate BC's (see~\cite{CGHST} for an extensive discussion of fermion BC's). The reason is that higher dimensional
fermions are necessarily non-chiral Dirac fermions. In a single flat extra dimension
a 5D fermion is given (in terms of 4D Weyl spinors) as
\begin{equation}
\left( \begin{array}{c} \chi (z)
\\[.2cm]
\bar{\psi} (z) \end{array} \right),
\end{equation}
where $\chi$ will denote a left-handed 4D Weyl fermion, and $\psi$ a right handed one. The equation of motion in
the bulk for such a field with bulk mass $m$ is given by
\begin{eqnarray}
-i \bar{\sigma}^{\mu} \partial_\mu \chi - \partial_{z_i} \bar{\psi} + m \bar{\psi} = 0,
\nonumber \\
-i \sigma^{\mu} \partial_\mu \bar{\psi} + \partial_{z_i} \chi + m  \chi = 0.
\end{eqnarray}
$\chi,\psi$ satisfy first order differential equations, so one must impose in total two BC's for the system.
The simplest BC's are $\psi|_{0,R'}=0$ (or $\chi|_{0,R'}=0$) which would imply the presence of a chiral
zero mode in $\chi$ with wave function $e^{-m z}$ (or for $\psi$ with wave function $e^{mz}$). Another possible
choice is that of twisted BC's, that is $\psi|_0=\chi|_{R'}=0$ (or the other way round) in which case there would
be no zero modes.

\begin{figure}[htb]
\begin{center}
\includegraphics[width=11cm]{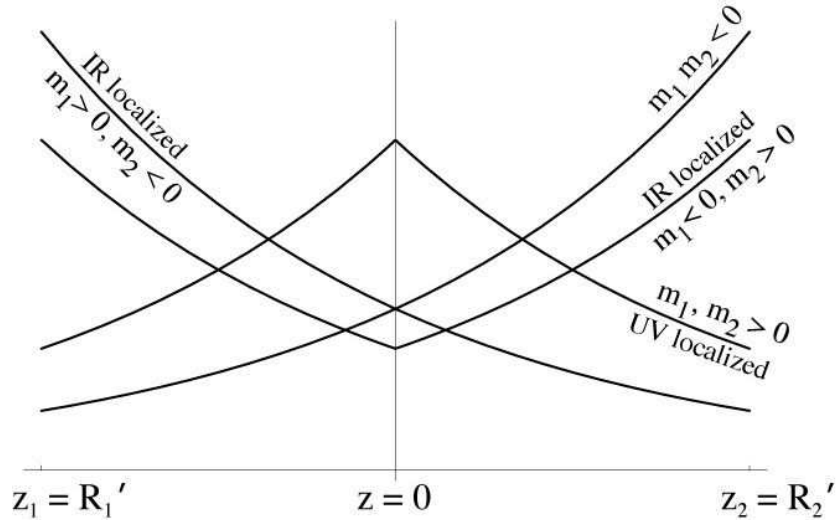}
\end{center}
\caption{The wave function of the fermions for various choices of the signs of the bulk masses. For the
continuous cases $m_1 m_2<0$, while for the case localized at the UV brane  $m_1,m_2>0$. The final
case (with localization at the two ends) corresponds to $m_1,m_2<0$. For all cases
we have chosen $|m_1|=|m_2|=1/R'_1=1/R'_2$.} \label{fig:fermions}
\end{figure}

Here we would like to construct a set-up with throats, which could give a theory with chiral zero modes, that could be
localized in either of the throats, or in both. For simplicity we will start with just two throats. If these were
not connected at the UV brane, there would be many possibilities for the zero modes. For example, we could have
a LH (or a RH) zero mode in each throat, or one LH in one and one RH in the other throat (assuming none of the fields
are twisted). However, we would like a case, when there is a single (for example LH) zero mode propagating in both
throats. This means that we need to start with a separate LH zero mode in both throats,
and remove one combination of them. Due to chirality this can only be done by adding a localized RH mode at the UV brane.
Consider for example the following action ($\overleftrightarrow{\partial_{z_i}}=
\overrightarrow{\partial_{z_i}}-\overleftarrow{\partial_{z_i}}$):
\begin{eqnarray}
S = && \int_0^{R'_1} d^4 x dz_1
\left(
- i \bar{\chi_1} \bar{\sigma}^{\mu} \partial_\mu \chi_1
- i \psi_1 \sigma^{\mu} \partial_\mu \bar{\psi}_1
+(\sfrac{1}{2}\,  \psi_1  \overleftrightarrow{\partial_{z_1}}  \chi_1
+ m_1 \psi_1 \chi_1 +{\rm h.c.} )
\right)+ \nonumber \\
&& \int_0^{R'_2} d^4 x dz_2
\left(
- i \bar{\chi}_2 \bar{\sigma}^{\mu} \partial_\mu \chi_2
- i \psi_2 \sigma^{\mu} \partial_\mu \bar{\psi}_2
+(\sfrac{1}{2}\, \psi_2  \overleftrightarrow{\partial_{z_2}}  \chi_2
+ m_2 \psi_2 \chi_2 +{\rm h.c.} )
\right)+ \nonumber \\
&& \int d^4x \left(m^{1/2}\,  \psi_0(\chi_1-\chi_2) +{\rm h.c.}\right)|_{z=0}
\end{eqnarray}
Here $1,2$ refer to two throats. We will also assume
that the BC in the absence of the UV brane would have been $\psi_1|_{0,R'_1}=\psi_2|_{0,R'_2}=0$. The
equation of motion with respect to the localized field will then enforce a modified BC
\begin{equation}
\chi_1|_{z=0}=\chi_2|_{z=0},
\end{equation}
while integrating the bulk equations of motion for the $\chi_{1,2}$ fields around the
UV brane ($\epsilon$ is a small distance away from the UV brane) imply
\begin{eqnarray}
{\bar{\psi}_1}{}_{|\epsilon} - {\bar{\psi}_1}{}_{|0} - m^{1/2}\, \psi_0 = 0 \\
{\bar{\psi}_2}{}_{|\epsilon} - {\bar{\psi}_2}{}_{|0} + m^{1/2}\, \psi_0 = 0.
\end{eqnarray}
Using the original BC at $z=0$, we are left with
\begin{equation}
\psi_1|_{z=\epsilon}=-\psi_2|_{z=\epsilon}.
\end{equation}
These new BC's will remove one of the combination of zero modes, and the remaining single left handed
zero mode will have the wave function
\begin{equation}
\chi_1(z)=A e^{-m_1z}, \chi_2 (z)=Ae^{-m_2z},
\end{equation}
and the remaining overall coefficient $A$ is determined by the
normalization of the kinetic term of this zero mode. If $m_1=m_2<0$ then the zero mode
will be localized at the two IR ends of the two throats. For $m_1=m_2>0$ the zero mode will be localized
around the UV brane in both throats, while for $m_1 m_2<0$ it will be localized at the end of the throat
with positive $m$. This is illustrated in Fig.~\ref{fig:fermions}.

We can easily generalize the above procedure for the case with arbitrary number $N$~throats. We will again
be interested in the case of a single chiral fermion (for example a left handed one)
propagating in all throats. Thus we need to start with a system that has a LH zero mode in all throats
(that is with BC's $\psi_i|_{0,R'_i}=0$). In order to remain with just one overall zero mode we need to remove
the additional zero modes via localized mass terms. One possibility would be to add Majorana masses, but we
are assuming that there are some conserved U(1) charges that generically forbid Majorana mass terms. Thus we
introduce $N-1$ localized RH modes $\xi_i$ at the UV brane, and add the localized Lagrangian
\beq
\int d^4x \sum_{i=1}^{N-1} \sum_{j=1}^N \left( m^{1/2}\, \tilde{\xi}_i \mathcal{K}_{ij} \chi_j + {\rm h.c.} \right)_{| z=0}\,.
\eeq
We will assume that the rank of the $N \times (N-1)$ matrix $\mathcal{K}$ is $N-1$, so that there is not any combination of the localized fields that can be decoupled, and only one zero mode is left in the spectrum.
The equations of motion of the localized fields, for the zero mode, will impose the following $N-1$ BC's at $z_i = 0$:
\beq
\sum_j \mathcal{K}_{ij} \chi_j = 0\,.
\eeq
The wave functions of the zero mode in the throats is
\beq
\chi_j = A_j e^{-m_j z_j}\,,
\eeq
where the coefficients $A_j$ are determined, up to an overall normalization, by the BC's on the UV brane.
Thus, the matrix $\mathcal{K}$ will determine in which throats the zero mode prefers to live, while, as in the 2-brane case, the signs and magnitudes of the bulk masses can yeld to a wide variety of different localizatrion patterns.

A simplified case, similar to the $N$ vibrating rods for a scalar field, is given by the following mass term:
\begin{equation}
\int d^4x \sum_{i=1}^{N-1}\left( m^{1/2}\, \xi_i (\chi_i-\chi_{i+1})+{\rm h.c.} \right)|_{z=0}.
\end{equation}
This will then imply the BC's $\chi_1=\chi_2= \ldots \chi_{N}$ at $z=0$, and the zero mode will be of the
form
\begin{equation}
\chi_i= A e^{-m_i z_i}, \ \  A= \left[ \sum_{i=1}^N \left( \frac{1-e^{-2 m_iR'_i}}{2 m_i}\right)\right]^{-\frac{1}{2}}.
\end{equation}

\section{Gauge Fields}
\label{sec:gauge}
\setcounter{equation}{0}
\setcounter{footnote}{0}
\label{gaugesection}

We now would like to extend the previous multi-throat set-up with
scalars  to allow gauge fields to also propagate  in the throats. The IR
BCs can be derived exactly as one would do for a
single throat theory so we only have to discuss the BC at the
UV brane. The bulk Lagrangians we consider are in flat space:
\begin{equation}
    \label{eq:gauge-bulk}
\mathcal{L} _{bulk,i}=- \frac{1}{4 g_{5,i}^2} F_{(i)M N}^{a\ 2}~.
\end{equation}
The results for the warped case are briefly summarized at the end of this section.

To find the BC's on the UV brane for the case where a single gauge field propagates in all the throats
let us start with  $N$~$SU(M)$ gauge
fields propagating in $N$~throats and consider how to describe in
field theory  the UV breaking of the $N$ gauge groups to the diagonal
$SU(M)$ gauge group. This can be achieved by introducing  $N-1$
bi-fundamental Higgs fields localized at the UV brane at $z_i=0$:
\begin{equation}
\mathcal{L}_{UV} = \sum_{\alpha=1}^{N-1}
\left( D_\mu \phi_\alpha^\dagger D^\mu \phi_\alpha ^{\vphantom{\dagger}}
- \lambda ({\rm Tr } \phi_\alpha^\dagger \phi_\alpha^{\vphantom{\dagger}} - c_\alpha^2)^2
\right)
\label{eq:gauge-higgs}
\end{equation}
where $\phi_\alpha$ transforms as a fundamental of $SU(M)_\alpha$
and an anti-fundamental of $SU(M)_{\alpha+1}$. The gauge covariant
derivative is given by ($T^{a}$ are the generators of the
gauge group in the fundamental representation and $A_{(\alpha)\, \mu}^a$ are the gauge fields in  
$\alpha^{th}$ throat)
\begin{equation}
D_\mu \phi_\alpha = \partial_\mu \phi_\alpha
- i  \left(A_{(\alpha)\, \mu}^a T^{a}  \phi_\alpha -\phi_\alpha T^{a}
A_{(\alpha+1)\, \mu}^a\right)
\end{equation}
As usual we expand the Higgs around its VEV:
\begin{equation}
\phi_\alpha = \frac{e^{i \pi_\alpha/f}}{\sqrt{2}} \left( v + h_\alpha \right)
\hspace{.6cm} \alpha=1\ldots N-1\,,
\end{equation}
where $\phi_\alpha$, $v$, and $h_\alpha$ are $M\times M$ matrices
and $\pi_\alpha=\pi^a_\alpha T^a$  and $f^2={\rm Tr}\,v^\dagger T^a
T^a v$. Since we are interested in breaking to the diagonal subgroup
we will take $v$ to be proportional to the identity. The Lagrangians
(\ref{eq:gauge-bulk})--(\ref{eq:gauge-higgs}) contain some bulk and
brane mixing terms involving $A_\mu$ that we want to cancel out with
a generalized $R_\xi$ gauge fixing term. Expanding up to quadratic
order:
\beq
\label{eq:lagrgaugesimp}
\mathcal{L}_{eff} &=&
\sum_{i=1}^N \int_0^{R'_{i}} \, dz_i\,
\left\{
\frac{1}{g_{5,i}^2} \left( -\frac{1}{4} F_{(i)\mu \nu}^{a\ 2}
+ \frac{1}{2} \left( \partial_{z_i} A^a_{(i)\mu} \right)^2
+  \frac{1}{2}
\left( \partial_\mu A^a_{(i)5} \right)^2- \partial_\mu A^a_{(i)5} \partial_{z_i} A_{(i)}^{a\ \mu} \right)\right\}
\nonumber \\
&&
+\sum_{\alpha=1}^{N-1} \frac{1}{2} {\rm Tr} \,\,v^\dagger
\left( \sfrac{1}{f} \partial^\mu \pi_\alpha -     A^{\mu}_{(\alpha)} +
A^{\mu}_{(\alpha+1)}  \right)
\left( \sfrac{1}{f} \partial_\mu \pi_\alpha -
A_{(\alpha)\mu} + A_{(\alpha+1)\mu} \right) v \nonumber \\
&&
+\sum_{\alpha=1}^{N-1} \frac{1}{2} {\rm Tr} \,\,\partial_\mu
h_\alpha^\dagger \partial^\mu h_\alpha -  \lambda ({\rm Tr} \,\, v^\dagger
h_\alpha)^2 + ...
\eeq
Integrating the derivative mixing term by parts gives:
\beq
\int_0^{R'_i} \,dz_i \,  A^a_{(i)5} \partial_{z_i} \partial_\mu A_{(i)}^{a\ \mu}
=
 - \int_0^{R'_i} \,dz_i\ \partial_{z_i}  A^a_{(i)5}  \partial_\mu A_{(i)}^{a\ \mu}
 +\left[ A^a_{(i)5} \partial_\mu A_{(i)}^{a\ \mu} \right]_0^{R'_i}\,.
\eeq
The mixing terms between vectors and scalars are then cancelled
by adding the bulk and the brane  gauge fixing Lagrangians (for
simplicity we will take all the brane gauge fixing parameters to be
equal to $\xi_b$):
\beq
\mathcal{L}_{GF}
&=&
-  \sum_{i=1}^N  \frac{1}{g_{5,i}^2} \frac{1}{2 \xi_i} \int \, dz_i
\left( \partial_\mu A_{(i)}^{a\ \mu} - \xi_i \partial_{z_i} A^a_{(i)5}  \right)^2
\nonumber \\
&&
-\sum_{i=1}^{N}
\left.
 \frac{1}{2 \xi_{b}}
\left( \partial_\mu A_{(i)}^{a\ \mu} + \xi_{b} \left( f  (\pi^a_{i}-\pi^a_{i-1}) -
\frac{A^a_{(i)5}}{g_{5,i}^2}  \right) \right)^2\right|_{z_i=0} \nonumber \\
&&
-\sum_{i=1}^{N}
\left.
 \frac{1}{2 \xi_{b}}
\left( \partial_\mu A_{(i)}^{a\ \mu} + \xi_{b}
\frac{A^a_{(i)5}}{g_{5,i}^2}  \right)^2\right|_{z_i=R'_i}\, ,
\eeq
(with the convention that $\pi^a_0=0$ and $\pi^a_N=0$). The
unitary gauge is realized in  the limit where all the $\xi$'s are
sent to infinity.   The full Lagrangian then leads
to the following equation of motion for $A_\mu$ ($q_i$
denotes the 4D momentum):
\begin{equation} \label{eom:Amu}
\partial_{z_i} \left( \partial_{z_i} A_{(i)\mu} \right) +
\left(q^2 \eta_{\mu\nu}+\left(\frac{1}{\xi_i}-1\right)q_\mu q_\nu\right)  A^\nu_{(i)}  =0\,,
\end{equation}
while the  BCs, fixed by requiring the vanishing of the boundary variation terms in
Eq.~(\ref{eq:lagrgaugesimp}), are:
\begin{equation}
\partial_{z_i} A^a_{(i)\mu} -g_{5,i}^2 f^2 \left(A^a_{(i) \mu }-A^a_{(i+1) \mu }\right)=0\,.
\label{eq:simpgaugebc}
\end{equation}
In the infinite VEV limit, the corresponding UV BCs are
\begin{eqnarray}
& \displaystyle
A^a_{(i)\, \mu} = A^a_{(i+1)\, \mu}\\
& \displaystyle \frac{\partial_{z_1} A^a_{(1)\, \mu}}{g_{5,1}^2}
\, + \ldots + \frac{ \partial_{z_N} A^a_{(N)\, \mu}}{g_{5,N}^2}\,
=0 \label{eq:masslessgaugeboson}
\end{eqnarray}
These BCs are analogous to case of scalar fields propagating in
multiple throats, thus the KK expansion of these fields will have
identical properties for the two cases.

Next we would like to discuss the question of under what
circumstances will there be physical modes in the scalar components.
The bulk equation of motion for the scalar fields $A_{(i)5}$ is:
\begin{equation}
    \label{eom:A5diag}
\xi_i \partial_z^2 A^a_{(i)5}  + q^2 A^a_{(i)5}  =  0 \ ,
\end{equation}
From this we see that the only physical modes that can survive in
the unitary gauge limit  $\xi\rightarrow \infty$ are those with
$q^2=0$.  Thus for theses modes we have
\begin{equation}
    \label{eom:A5physical}
 \partial_z^2  A^a_{(i)5}   =  0 \ .
\end{equation}
The BC's for the $\pi$'s and $A_{(i)5}$'s are obtained by requiring that the variation of the action vanishes at the UV brane:
\beq
\delta \mathcal{L}_{scalar}
&=&
-\sum_{\alpha=1}^{N-1}
\partial_\mu \partial^\mu \pi^a_\alpha  \delta \pi^a_\alpha
+\sum_{i=1}^{N} \frac{\xi_i}{2g_{5,i}^2} \delta A^a_{(i)5} \partial_{z_i}  A^a_{(i)5}|_{z_i=0}\\
&&
-\sum_{i=1}^{N} \left.
  \xi_{b}
   \left( f  (\pi^a_{i}-\pi^a_{i-1}) -
\frac{A^a_{(i)5} }{g_{5,i}^2} \right) \right|_{z_i=0}
\left( f  (\delta \pi^a_{i}- \delta \pi^a_{i-1}) -
\frac{\delta A^a_{(i)5} }{g_{5,i}^2} \right)
\eeq
In the unitary gauge the terms in the $A_5$ and $\pi$ equations of
motion that are multiplied  by gauge parameters must vanish.  So we
see that $\pi^a_1$ and $\pi^a_{N-1}$ are fixed in terms of
$A_{(1)5}^a$ and $A_{(N)5}^a$.  The remaining $N-2$ equations from
$\xi_b$ terms fix the remaining $\pi^a_i$'s and require
\beq
0=\sum_{i=1}^{N} \left. \frac{A^a_{(i)5}}{g_{5,i}^2} \right|_{z_i=0}
\label{eq:A5sum}
\eeq
By requiring that the terms proportional to $\xi_i$ vanish, we also obtain
\beq
\partial_{z_i} A^a_{(i)5}|_{z_i=0} = 0
\label{eq:A5der}
\eeq
We can
see that as expected, the BCs for the $A_5$
components are the same as for the $A_\mu$ components, except that the
Dirichlet BC is replaced by a Neumann BC and vice versa. The BCs
(\ref{eq:A5sum})--(\ref{eq:A5der}) together with the bulk equation of motion implies
that for any physical mode the $A_{(i)5}$'s have to be constant in
every throat, and due to the BC (\ref{eq:A5sum}) these constants
have to add up to zero. Thus there can at most be $N-1$ independent
scalar modes. How many of these $N-1$ are actually physical will
depend on the BCs at the IR branes. If there is no additional
symmetry breaking on any brane, or if there is symmetry breaking on
at most one brane, then none of the scalar $A_{(i)5}$ modes will be
physical. The reason is that the BC on a given IR brane for
$A_{(i)5}$ will be Neumann $\partial_{z_i} A^a_{(i)5}|_{z_i=R'}=0$
for the case when a particular generator $a$ is broken by a VEV of a
scalar field on the brane, while it will be Dirichlet
$A^a_{(i)5}|_{z_i=R'}=0$ for the case there is no symmetry breaking.
If there is a Dirichlet BC at the bottom of at least $N-1$ throats
then at least $N-1$ of the $A_{(i)5}$'s are set to zero, and since
all $N$~have to also add up to zero, we see that they all have to vanish. So there is no
physical mode left. Thus in order for a physical scalar zero mode to
exist in any particular gauge direction, the generator corresponding
to that direction has to be broken at least  twice (on the bottom
of at least two throats). This also immediately implies, that if a
particular generator is broken on $k$ throats, there will be $k-1$
physical scalar zero modes appearing.

The counting of the physical
scalar modes has a very simple holographic interpretation (at least
for the case when all the throats are warped). If the extra
dimensions were warped, then the set-up we have produced would
correspond to  $N$~different 4D CFT's (corresponding to the bulks
of the warped throats), each of which has an $SU(M)$ global
symmetry. The set-up around the UV brane ensures that only the
diagonal subgroup of these global symmetries is actually gauged.
Then if the symmetry is broken on the bottom of $k$ throats that
will correspond to $k$ of the CFT's spontaneously breaking its own
global symmetry, each of which will produce a Goldstone boson.
However, since the diagonal subgroup is gauged, one combination of
these Goldstone bosons is eaten, leaving us with $k-1$ physical
massless Goldstone modes. We have seen that the BC's derived here
exactly reproduce this expected answer. Even though we have
only considered the flat throat case here, the generalization to the warped case is straightforward.
With bulk Lagrangians of the form
\begin{equation}
    \label{eq:gauge-bulkapp}
\mathcal{L} _{bulk,i}=- \mathcal{K}_{(i)}(z_i) \frac{1}{4 g_{5,i}^2} F_{(i)M N}^{a\ 2}~,
\end{equation}
following the same procedure as before, we obtain the following BC's for the $A_\mu$ components
in the infinite VEV limit
\begin{eqnarray}
& \displaystyle
A^a_{(i)\, \mu} = A^a_{(i+1)\, \mu}\\
& \displaystyle
\frac{\mathcal{K}_{(1)}(z)}{g_{5,1}^2} \, \partial_{z_1} A^a_{(1)\, \mu}
+ \ldots + \frac{\mathcal{K}_{(N)}(z)}{g_{5,N}^2}\,  \partial_{z_N} A^a_{(N)\, \mu} =0
\label{eq:warpedmasslessgaugeboson}
\end{eqnarray}
while the BC's for the $A_5$ components read
\beq
0=\sum_{i=1}^{N} \left.
\frac{\mathcal{K}_{(i)}}{g_{5,i}^2} A^a_{(i)5} \right|_{z_i=R} \\
0= \partial_{z_i} \left(\mathcal{K}_{(i)} A^a_{(i)5}\right)|_{z_i=R}
\eeq

\section{Gravity}
\label{sec:gravity}
\setcounter{equation}{0}
\setcounter{footnote}{0}

Let us finally discuss how an appropriate gravitational background
and fluctuations could be obtained in these throats. For useful
references for this section see~\cite{RSrefs}. For the background
metric we will be assuming that it is conformally flat in every
throat and is given by
\begin{equation}
g_{MN}^{(i)}= e^{-A_i(z_i)} (dx^\mu dx^\nu \eta_{\mu\nu} -dz_i^2),
\end{equation}
where the coordinates $z_i$ run from the UV brane $z_i=0$ to the
individual IR branes $z_i=R_i'$. We will assume that the metric is
continuous at the UV brane, that is
\begin{equation}
A_i=A_{i+1}=0.
\end{equation}
Here we have  rescaled the overall warp factor on the UV brane to be
equal to one. If there are different cosmological constants
$\Lambda_i$ in the different throats, then the warp factors will be
given by
\begin{equation}
A_i(z_i)=2 \log ( z_i/R_i +1),
\end{equation}
where the AdS curvatures in the different throats are given by their
usual expressions
\begin{equation}
1/R_i^2=-\frac{\kappa^2 \Lambda_i}{6} ,
\end{equation}
and $\kappa^2$ is the 5D Newton constant (related to the 5D Planck
scale by $\kappa^2=1/(2 M_*^3)$). These will completely fix the
solutions, however one still has to satisfy the generalized Israel
junction condition~\cite{Israel}. In this case it will be given by 
\begin{equation}
\sum_i \partial_{z_i} A_i{}_{|z_i=0}=\frac{2 \kappa^2}{3} T,
\end{equation}
where $T$ is the tension of the junction (i.e. the UV brane). This will just result in
the usual Randall-Sundrum (RS)-type fine tuning between the brane tensions and the
various throat cosmological constants:
\begin{equation}
\sum_i \frac{1}{R_i} =\frac{\kappa^2}{3} T.
\end{equation}

One can also find the graviton KK modes easily in this background.
Assuming that the metric fluctuations are of the form
\begin{equation}
ds_i^2=e^{-A_i(z_i)} (dx^\mu dx^\nu
(\eta_{\mu\nu}+h^{(i)}_{\mu\nu}(x,z_i))-dz_i^2)
\end{equation}
and using transverse traceless gauge (that is neglecting the
radion-type scalar solutions) $h^\mu_\nu=\partial^\mu h_{\mu\nu}=0$
we find that the Einstein equations in the throats give the usual
RS-type equations:
\begin{equation}
(\partial^\mu \partial_\mu-\partial_{z_i}^2)
h_{\mu\nu}^{(i)}+\frac{3}{2} \partial_{z_i} A_i \partial_{z_i}
h_{\mu\nu}^{(i)}=0. \label{eq:bulkgrav}
\end{equation}
All the terms including the brane tension are compensated by the
effects of the background metric, so the generalized Israel junction
condition at the UV brane for the perturbations will just be:
\begin{equation}
\sum_i \partial_{z_i} h_{\mu\nu}^{(i)} =0,
\end{equation}
while the metric should still be continuous on the UV brane
requiring
\begin{equation}
h_{\mu\nu}^{(i)}=h_{\mu\nu}^{(i+1)}.
\end{equation}
Thus the BC's for the graviton modes are completely analogous to
the case of scalars and gauge bosons with a single zero mode. For the particular case of a
graviton zero mode in a warped throat (\`a la RS), we can easily solve these equations, since the
solution of (\ref{eq:bulkgrav}) are just constant in every throat,
with the constants equal at the UV brane. The overall normalization
would then be fixed by calculating the kinetic term of the graviton
zero mode.
\section{Applications}
\label{sec:applications}
\setcounter{equation}{0}
\setcounter{footnote}{0}

\subsection{A throat for every family}

\begin{figure}[htb]
\begin{center}
\includegraphics[width=7cm]{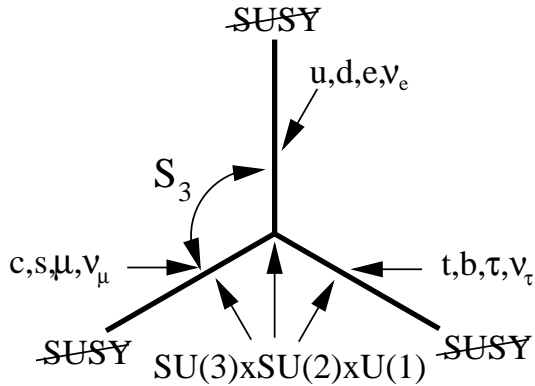}
\end{center}
\caption{Illustration of the throat construction of the
model with a throat for every family.} \label{threefamilies}
\end{figure}

One obvious application of the multi-throat construction is to have
a set-up where the gauge bosons propagate on all the throats but a
family of quarks and leptons  lives in a single throat. With three
throats, we can have  the three families each propagating in
different throats. In order to be able to generate mixings among
different families the Higgs field has to have an overlap with the
UV brane. Thus one would expect that in order to solve the gauge
hierarchy problem one would have to make the setup supersymmetric.
The supersymmetric flavor problem could then be resolved
geometrically, if one assumes that the three throats were identical
(that is they are related by an $S_3$ exchange symmetry). In this case 
one could
imagine that the only source of supersymmetry breaking was due to
strong dynamics at the bottom of the throats giving rise to equal
soft supersymmetry breaking terms localized on the IR branes. Since
we are assuming that the three throats are identical, one will get
equal soft breaking terms for the three families, and since they are
separated in different throats they will be flavor diagonal. In
order for the Higgs to pick up soft breaking terms it has to then
propagate in the three throats. Since we are assuming that the 
three throats are identical this will also imply that the Yukawa couplings in 
the throats are equal for the three families. A hierarchy among the three 
families can then be obtained by adding terms to the UV brane such that 
the VEV's of the Higgses in the three different throats will be the largest
for the throat containing the third generation, and the smallest for the first
generation. The mixings among the generations and the differences of 
masses within a generation can then be adjusted by adding some additional 
UV-brane localized Yukawa couplings. The setup is illustrated in
Fig.~\ref{threefamilies}.

\subsection{A GUT model for doublet-triplet splitting}

One can embed this  scenario in a model of doublet-triplet splitting in SUSY GUTs along the lines
of~\cite{BDS,WittenGUT}. Thus these GUT models  can be given a simple geometric
interpretation. The essence of the model in~\cite{BDS} is to have
three SU(5) groups broken by various bifundamental fields $Z_i$ and
$\bar{Z}_i$ to the diagonal SU(3)$\times$ SU(2)$\times$U(1)
subgroup. A key element is the $S_3$ exchange symmetry of the three
SU(5) factors and an additional $Z_2$ symmetry that will forbid
direct mass terms of the Higgs fields.

\begin{figure}[htb]
\begin{center}
\includegraphics[width=7cm]{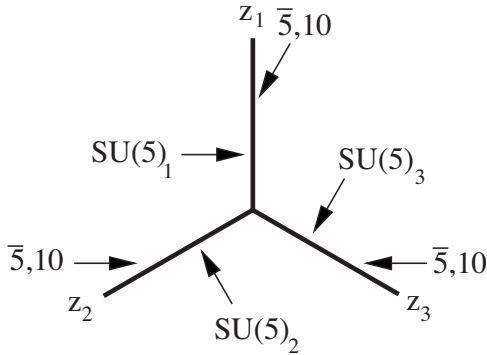}
\end{center}
\caption{Illustration of the throat construction of the
$SU(5)^3$ model.} \label{doubletriplet}
\end{figure}

The geometric construction is again based on a supersymmetric
three-throat setup, each of which have an SU(5) group propagating in
them. The three throats have again an $S_3$ exchange symmetry as in Fig.
\ref{doubletriplet} . We
assume that half of the eight supercharges are broken on the IR
brane via BCs, and thus only an ${\cal N}=1$ set of
fields have light modes. On the UV brane we introduce the
bifundamental fields $Z_i$, $\bar{Z}_i$ and a generic superpotential
(as in \cite{BDS})
\begin{equation}
W(Z)_{UV}= M_1 \sum_j (Z_j \bar{Z}_j) +\frac{1}{M_2} \sum_{j\neq k} (Z_j
\bar{Z}_j) (Z_k \bar{Z}_k) +\ldots ,
\end{equation}
where the extra terms correspond to additional dimension 
four superpotential terms. Among the possible
set of supersymmetric vacua are those with VEVs
\beq
&&\langle Z_1 \rangle =\langle \bar{Z}_1 \rangle = v_1 \ {\rm
diag}(1,1,1,0,0), \\ &&\langle Z_2 \rangle =\langle \bar{Z}_2
\rangle=\langle Z_3 \rangle =\langle \bar{Z}_3 \rangle = v_2 \ {\rm
diag}(1,1,1,1,1).
\eeq
With these VEVs the three SU(5)'s in the three throats are broken to
just the SM groups. There are six separate Higgs fields
$H_i,\bar{H}_i$ (which are originally six hypermultiplets, but due to
the BC's on the IR brane only one chiral component will have a zero
mode). In order to forbid the direct mass term for these fields one
needs to assume that there is a $Z_2$ symmetry $\{ \bar{H},Z,\bar{Z}
\} \to -\{ \bar{H},Z,\bar{Z} \}$ which will only allow the
superpotential coupling on the UV brane
\begin{equation}
W(H,Z)_{\rm trilinear}= \lambda \sum_{i,k} H_i Z_{ik} \bar{H}_k,
\end{equation}
where we have defined $Z_{ik}=Z_j$ for $j\neq i,k$. With this
additional UV brane superpotential all components of the Higgs
fields $H_i,\bar{H}_i$ will get a mass of order $v_{1,2}$, except
for one pair of SU(2) doublets, which will be identified with the
MSSM Higgs doublets. This will also automatically solve the
$\mu$-problem, and an order $M_{weak}$ $\mu$-term can be generated
if after minimizing the potential for the $Z,\bar{Z}$-fields the
VEV's will get shifted due to the addition of the soft breaking
terms. The model of \cite{WittenGUT} can be implemented similarly,
with the added convenience that the UV brane VEV's of the
$Z,\bar{Z}$ can now be interpreted as Wilson lines along the
coordinates parameterizing the direction along the UV head.

\subsection{A simple geometric construction for trinification}

Quite a lot of effort has been devoted~\cite{EDtrin} to naturally implement
a grand unified trinification~\cite{trin} model
based on the gauge group $SU(3)^3/Z_3$ into extra dimensions. The multi-throat
construction clearly offers a simple implementation for such a theory.
In 4D the gauge group is given by $SU(3)_c\times SU(3)_L\times SU(3)_R$,
with $SU(3)_c$ being ordinary QCD, $SU(2)_L \subset SU(3)_L$, and
$Y=-\frac{1}{\sqrt{3}}(T^8_L+\sqrt{3} T^3_R+T^8_R)$. One family
of SM matter can be embedded into the representation $27=(3,\bar{3},1)+
(1,3,\bar{3})+(\bar{3},1,3)$. This contains the usual SM fermions,
plus some additional massive vector-like fields: an additional $SU(2)_L$
singlet vector-like bottom, an additional vector-like $SU(2)_L$ doublet
(color singlet) plus two right handed neutrinos.

The symmetry breaking is obtained by at least two Higgs fields in the
same representation~27. Clearly, since QCD is unbroken only the $(1,3,\bar{3})$
component of these higgses can get a VEV, and the assumption is that one of
these higgses $\Phi$ will get a VEV of the form
\begin{equation}
\langle \Phi \rangle =\left( \begin{array}{ccc} ~&~&~\\ &&\\ &&v \end{array}
\right)
\end{equation}
breaking $SU(3)_L\times SU(3)_R$ to $SU(2)_L\times SU(2)_R\times U(1)_{B-L}$,
while the other Higgs $\chi$ has a VEV of the form
\begin{equation}
\langle \chi \rangle =\left( \begin{array}{ccc} ~&~&~\\ &&\\ &v'& \end{array}
\right)
\end{equation}
breaking the group to a different $3-2-2-1$ subgroup. In total the two VEV's
would leave 3-2-1 invariant. Electroweak symmetry breaking is achieved 
if further (electrically and color neutral) components of the Higgs fields
get VEV's of the order of the electroweak scale.

The extra dimensional construction based on throats would work as follows:
fermions are bi-fundamentals under the SU(3) groups, and assuming they
live in separate throats, the individual $SU(3)$ gauge bosons have to be
propagating in two separate throats each:

\begin{equation}
SU(3)_c \to  (z_1,z_2), \ \ SU(3)_L\to  (z_2,z_3), \ \ SU(3)_R\to (z_1,z_3).
\end{equation}
This can be achieved using a construction for the gauge fields of the
sort presented in Sec.~\ref{sec:gauge}. Fermions are then localized in single
throats, obviously with the above assignment of the gauge groups we would
find
\begin{equation}
(3,\bar{3},1)\to z_2, \ \ (1,3,\bar{3}) \to z_3, \ \ (\bar{3},1,3) \to
z_1.
\end{equation}

\begin{figure}[htb]
\begin{center}
\includegraphics[width=7cm]{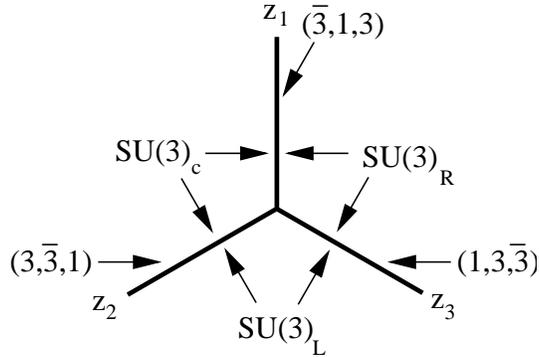}
\end{center}
\caption{Illustration of the throat construction of the
model for trinification.} \label{trinif}
\end{figure}

This is illustrated in Fig.~\ref{trinif}.
Since the symmetry breaking should be felt by all the fermions, the Higgs
fields should be localized on the Planck brane. Thus in order to
explain the hierarchy the theory should be supersymmetric.
Fermion hierarchies
could then be explained by varying the localization parameters of the fermions.

\section{Conclusions}
\label{sec:conclu}
\setcounter{equation}{0}
\setcounter{footnote}{0}

We have shown how a general theory on a manifold with many throats
can be described in a field theory limit as a collection of five
dimensional warped spaces intersecting on a single UV brane. We have
also shown how the Lagrangian on the UV brane determines the
BCs for the extra dimensional fields, giving
examples with scalars, gauge bosons, and fermions. This set-up
provides a whole new set of tools for model builders to exploit. We
have shown the appropriate BCs for scalar, fermion, gauge boson and
graviton fields. We have provided three very simple sample
applications of these ideas. In the first the three generations of
the MSSM propagate in separate throats, and the supersymmetric
flavor problem is solved via the exchange symmetry of the three
throats. The second example we show how  this scenario can be embedded in a beautiful model of
doublet-triplet splitting in SUSY GUTs which can be also interpreted
geometrically. The third example provides a geometric implementation of
trinification.


\section*{Acknowledgements}
We thank Nemanja Kaloper and John March-Russel for collaborations at the early stage of this project.
We thank Hsin-Chia Cheng, Jack Gunion, Ryuichiro Kitano, Yasunori Nomura, and Michael Peskin for useful discussions and comments.
The research of G.C. and C.C.
is supported in part by the DOE OJI grant DE-FG02-01ER41206 and in part
by the NSF grants PHY-0139738  and PHY-0098631.
C.G. is supported in part by the RTN European Program
MRTN-CT-2004-503369 and by the ACI Jeunes Chercheurs 2068.
J.T. is supported by the US Department of Energy
under grant  DE-FG02-91ER40674. This collaboration is partly supported by the CNRS/USA grant 3503.
We also thank the Aspen Center for Physics for its hospitality
while part of this project has been performed.

\end{document}